\documentclass[aps,preprint]{revtex4}
\usepackage[dvips]{color}
\usepackage{graphicx}
\usepackage{newlfont}

\begin{document}
\title{Low Loss Superconducting Titanium Nitride Coplanar Waveguide Resonators}
\author{M. R. Vissers$^{a}$, J. Gao$^{a}$, D. S. Wisbey$^{a}$, D. A. Hite$^{a}$, C.C. Tsuei$^{b}$, A.D. Corcoles$^{b}$,  M. Steffen$^{b}$, and D. P. Pappas$^{a}$}
\affiliation{${a})$National Institute of Standards and Technology, Boulder, CO\\
  ${b})$IBM T. J. Watson Research Center, Yorktown Heights, NY.\\
}

\begin{abstract}
Thin films of TiN were sputter-deposited onto Si and sapphire wafers with and without SiN buffer layers.  The films were fabricated into RF coplanar waveguide resonators, and internal quality factor measurements were taken at millikelvin temperatures in both the many photon and single photon limits, i.e. high and low power regimes, respectively.  At high power, internal quality factors ($Q_i$'s) higher than $10^7$ were measured for TiN with predominantly a (200)-TiN orientation. Films that showed significant (111)-TiN texture invariably had much lower $Q_i$'s, on the order of $10^5$. Our studies show that the (200)-TiN is favored for growth at high temperature on either bare Si or SiN buffer layers. However, growth on bare sapphire or Si(100) at low temperature resulted in primarily a (111)-TiN orientation. Ellipsometry and Auger measurements indicate that the (200)-TiN growth on the bare Si substrates is correlated with the formation of a thin, $\approx 2$ nm, layer of SiN during the pre-deposition procedure.  In the single photon regime, $Q_i$ of these films exceeded $8\times10^5$, while thicker SiN buffer layers led to reduced $Q_i$'s at low power.\\
\\
Contribution of U.S. Government, not subject to copyright.
\end{abstract}
\maketitle

The quest for materials that have low loss in RF resonant structures at low temperatures is an area of great interest for quantum computation \cite{Wallraff,OConnell} and photon detection \cite{Day}. Low loss, i.e. high quality factor, in these applications is necessary to have long resonant lifetimes at low power, and well resolved frequencies and low noise at high power. Important examples are the storage of arbitrary quantum states of single photons in superconducting resonators \cite{Hofheinz} and multiplexed readout of kinetic inductance photon detectors \cite{Mazin,Baselmans}.

Presently, superconducting materials such as Al, Re, and Nb on crystalline substrates such as silicon and sapphire are capable of producing low loss, $\delta_i$, and therefore high internal quality factors ($Q_i={1 \over \delta_i}$) in the $10^5 - 10^6$ range in the many-photon regime \cite{WangQC,Gao,Barends}. However, when restricted to the single photon power levels used in superconducting quantum information applications, quality factors are reduced to the $10^4 -10^5$ range. This limits reproducible lifetimes, $\tau = {Q \over 2 \pi f}$, to be on the order of a microsecond when operated in the 1 - 10 GHz range. It is well accepted that extraneous two-level systems (TLS) in oxides at surfaces, interfaces, and dielectrics contribute predominantly to the losses in these structures \cite{WangQC,Gao,Lindstrom,Barends,You,Steffen,Macha}. It is therefore reasonable to expect that a material that has a lower tendency to oxidize would result in lower loss.  Nitride surfaces have been known to be very stable, especially against oxidation, and resonators based on NbTiN and TiN, for example,  have shown promise in superconducting resonators \cite{Barends, Leduc}.

In particular, Leduc, {\it et al.}, have recently shown that resonators with exceptionally high $Q_i$, on the order of $10^7$ at high power, can be fabricated from TiN grown on hydrogen terminated, intrinsic silicon (H:i-Si)\cite{Leduc}. TiN is an example of a hard, stable material that is widely used as a coating for the machining industry, diffusion barriers for semiconductors, and in MEMS devices \cite{Watmon, Sidhwa, Weon}. It is also a superconductor with a relatively high $T_C$, with stoichiometric TiN yielding higher than 4 K \cite{Spengler}. However, the growth mode and morphology of TiN is sensitive to a wide range of parameters, including film thickness, substrate temperature, bias, and crystallinity \cite{Oh,Ho,Sundgren,Igasaki,Kiuchi,Harra,Patsalas}. In addition, results of nitride-based devices fabricated on sapphire, a traditionally low loss substrate, have shown relatively high loss results \cite{Andreone, Zmuidzinas}. In this work we show that high $Q_i$ at high power is correlated with the (200)-TiN texture. This orientation is stabilized with a thin buffer layer of SiN. In addition, we are also interested in the behavior of these materials at low power for applications in the field of quantum computing, and we find that devices with only the 2nm SiN buffer layer formed during the pre-sputtering procedure show $Q_i$ up to $8\times10^5$ in the single photon regime.

In general, to improve the electrical characteristics of films, especially in the RF region, it is reasonable to work in the direction of reducing stress and increasing the density of the superconducting material. From this perspective, we took the approach of growing the TiN oriented with the plane of lowest surface energy, i.e. the (200) face for the NaCl structure. From the literature, it is known that to stabilize high density (200)-TiN, films should be grown at high temperature with a substrate bias, relatively thin, and on an amorphous surface \cite{Patsalas,Igasaki,Oh}.

Our TiN films, 40 nm thick, were reactively DC sputter deposited onto c-plane sapphire and H-terminated, intrinsic Si(100), (H:i-Si), substrates ($>$15 k$\Omega \cdot$ cm) with and without pre-deposited SiN buffer layers. The sapphire was prepared with an {\it in situ} anneal to 500 $^\circ$C, while the Si wafers were prepared by etching in a 10:1 H$_2$O:HF solution to remove any native oxide and to hydrogen terminate the surface.  {\it In situ} ellipsometry and Auger data on the freshly loaded H:i-Si show evidence of about one monolayer (0.2 nm) surface oxide, compared to $2 - 3$ nm for unetched wafers.  However, there is typically a 1 minute soak while the sputtering source is operating before the shutter is open. We believe that this is important because it creates the opportunity to form a thin nitride prior to TiN deposition if the Si substrate is hot. This is discussed more in-depth below. The TiN deposition was performed at 500 $^\circ$C with a DC bias on the substrate of -100 V. The pressure was held at 5 mTorr in a reactive mixture of 3:2  argon to nitrogen, with a growth rate of 2 nm/min. Our films are in the high nitrogen percentage limit, and we measured $T_C$'s between 4.2 to 4.6 K, in agreement with previous studies \cite{Spengler}.  {\it In Situ} RHEED indicates that the films grown on sapphire are well ordered and crystalline, while those grown on Si and SiN are highly disordered and polycrystalline.  Finally, AFM studies of the surface indicate the RMS roughness of the films is typically less than 1 nm.

For the RF loss studies, the TiN films were patterned into frequency multiplexed, coplanar waveguide (CPW), half-wave resonators capacitively coupled to a microwave feedline \cite{Day}. This arrangement permits the extraction of $Q_i$ from the $S_{21}$ transmission measurement \cite{MazinThesis,JGaoThesis}. The CPW resonators had a 3 $\mu$m wide centerline and a 2 $\mu$m gap. They were patterned from the TiN film using standard photolithography techniques and a reactive ion etch (RIE) in an SF$_6$ plasma.  The resonances were measured in an adiabatic demagnetization refrigerator at temperatures below 100 mK.  The sample box holding the resonator chip was magnetically shielded with an outer cryoperm and inner superconducting shield.  Measurements were performed using a vector network analyzer with a combination of attenuators (room temperature and cold) on the input line to achieve the appropriate power level at the device input port and a microwave isolator and HEMT amplifier on the output. The power in the resonators was calculated in terms of the electric field, E, in the standard manner from the attenuation, measured resonance parameters and the CPW gap \cite{JGaoThesis}.

\begin{figure}[ht]
\centerline{\includegraphics[scale=0.6]{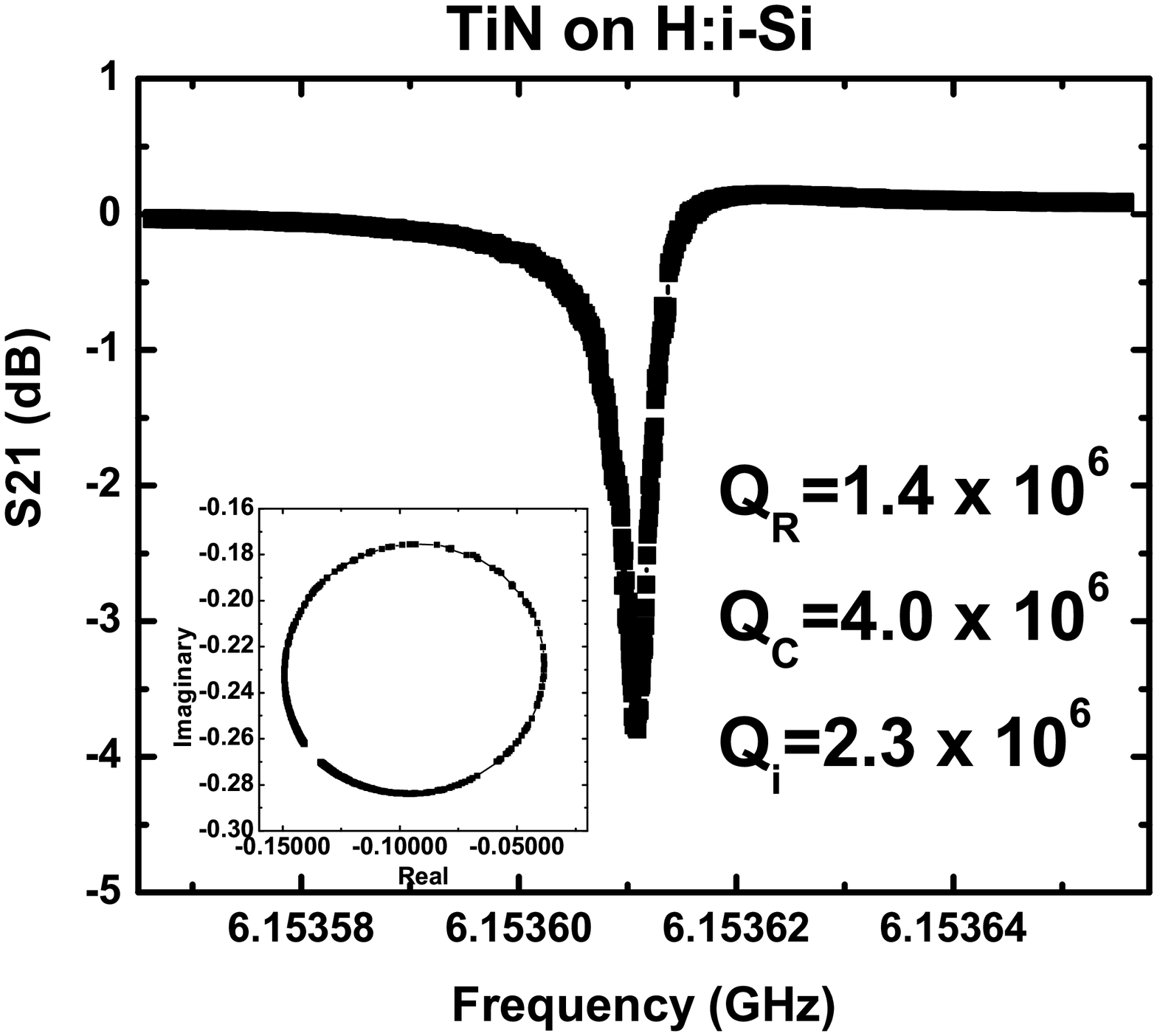}}
\caption{Measured S21 vs frequency trace  of a TiN resonator in the many photon limit.  The resonance is characterized by a dip in the magnitude and a circle in the complex plane (inset).   The resonator is 3um wide with a 2um gap.  The temperature is 75mK.  Note that the resonant, coupling and internal quality factors all exceed 1 Million.}
  \label{Resonance}
\end{figure}

The low loss of the TiN necessitates very weak coupling (high coupling $Q_C$) between the resonator and the feed line for accurate measurement of $Q_i$. Therefore, $Q_C$'s ranging from 500k to 5M were used in the resonator design. The measured $Q_i$'s of the different designs were not dependent on the coupling $Q_C$'s. Figure \ref{Resonance} shows a typical resonance measured at high power. We find measured total $Q_R$'s as well as $Q_i$'s well in excess of 1 million. Some of the best resonators in our studies show $Q_i$'s higher than $10^7$, an order of magnitude higher than typical Nb, Al, or Re devices made in this geometry, and in agreement with other high power measurements of nitride-based devices \cite{Leduc}.  Since any TLS are fully saturated in the many photon regime, the high power measurements are an indication that there is low intrinsic loss of the superconducting TiN surface \cite{VanDuzer}.

\begin{figure}[ht]
\centerline{\includegraphics[scale=0.45]{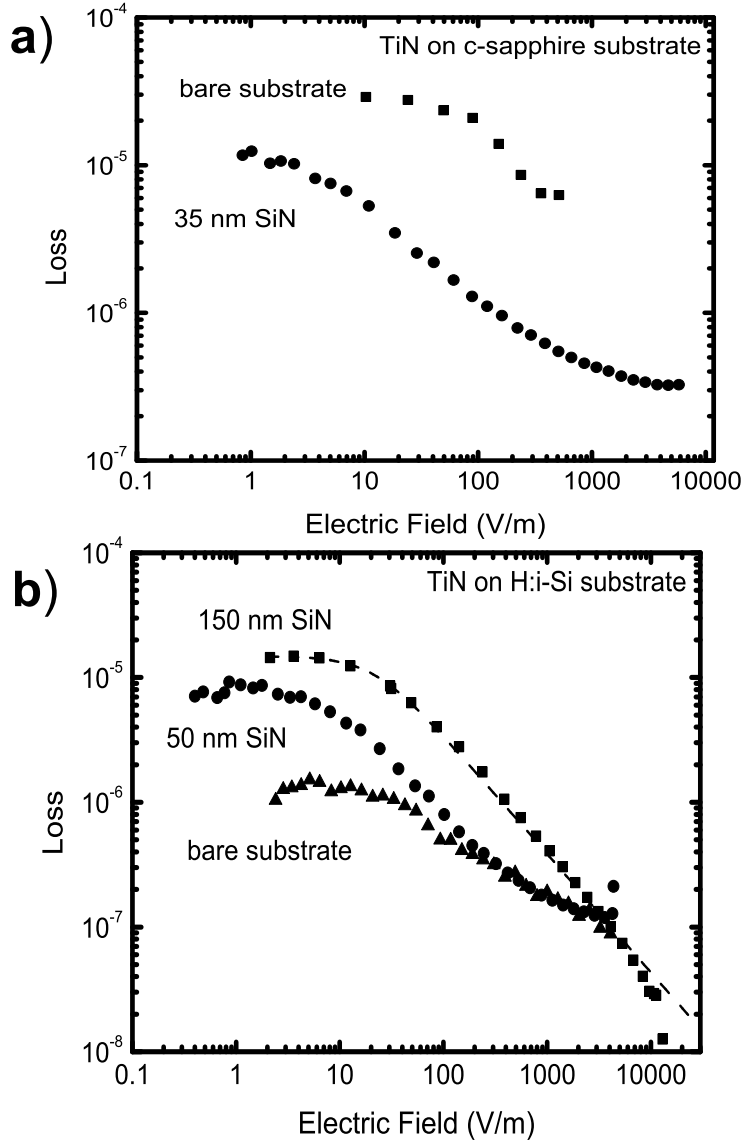}}
\caption{Internal quality factor, $Q_i$, of TiN films grown on sapphire, SiN and Si surfaces.  The TiN on sapphire has the lowest Q for both high and low powers.  The TiN on both SiN samples as well as Si has similar quality factor in the many photon limit, but at lower powers the increased participation of the higher loss  SiN contributes to increased loss at low power.  In all cases the loss saturates at low power in the single photon limit.}
\label{QvE}
\end{figure}

Figure \ref{QvE} shows loss ($1/Q_i$) as a function of power (in terms of electric field) for a number of samples. Most significantly, we find that the TiN grown on the bare H:i-Si substrate (Figure \ref{QvE}(b)) has nearly two orders of magnitude lower loss than on bare sapphire (Figure \ref{QvE}(a)) for films with growth conditions nominally the same. From the observation that the films grown on sapphire are more crystalline, we chose to use an amorphous buffer layer on the sapphire to inhibit nucleation of non-equilibrium epitaxy, thus allowing the TiN to grow in its low-energy orientation. SiN was chosen because it has lower loss than SiO$_X$ \cite{Paik}. As seen in Figure \ref{QvE}(a), for  35 nm SiN on sapphire, we recover the very low loss behavior at high powers seen on Si substrates. The residual loss is in line with what is expected for the filling factor and TLS contribution of the SiN.   To make a more direct comparison, we also used SiN buffer layers on H:i-Si substrates. The loss curves from TiN on 50 nm and 150 nm SiN buffer layers show the same low loss at high power as for TiN on the H:i-Si. The magnitude of the low power loss is in qualitative agreement with the thickness of the SiN on the substrate, with the 150 nm buffer layer sample showing the highest loss by a factor of 2-3, as expected. We fit to the data for TiN/150 nm SiN, where the low power loss is much higher than the background high power loss. This allows for a reliable fit to the expected TLS loss, $\delta_{TLS}$ power dependence, i.e.
\begin{equation}
\delta_{TLS}(E,T)={\delta_{TLS}(E<<E_C,T=0)tanh({\hbar\omega \over 2k_BT}) \over ({1+(E/E_C)^\Delta})^{1/2}},
\end{equation}
where $E_C$ is the critical field for saturation of the TLS and the exponent, $\Delta$, should be equal to 2 for true TLS loss \cite{JGaoThesis}. From our fit, we find $\Delta=1.9\pm{0.1}$, showing that the extra loss in the SiN buffer layers is consistent with TLS theory.

\begin{figure}[ht]
\centerline{\includegraphics[scale=.6]{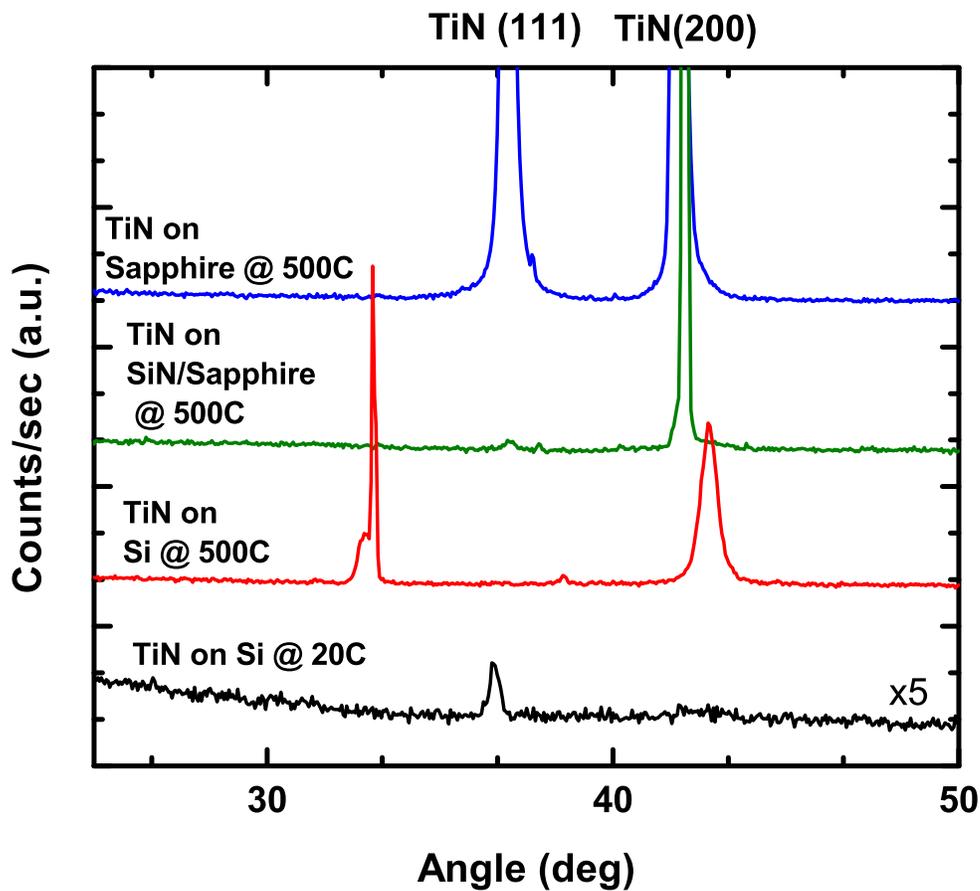}}
\caption{$\theta-2\theta$ XRD scans of TiN films on sapphire, Si and SiN/Si at 500 $^\circ$C, and Si at 20 $^\circ$C.  The (111)-TiN peak at $2\theta=36^\circ$ is present on the sapphire substrate as well as the for Si room temperature.  Both of these films exhibited low internal Q at high and low power.  The TiN grown at high temperature on Si and SiN both exhibit primarily (200)-TiN peak at $2\theta$ around $42^\circ$. The sharp peak at $33 ^\circ$ on the high temperature TiN on Si is due to the XRD being performed on a patterend sample with exposed Si regions. }
\label{xrayV3}
\end{figure}

To better understand the correlation of the RF properties and film structure, we conducted {\it ex situ} x-ray diffraction (XRD) and scanning electron microscopy (SEM) as well as {\it in situ} ellipsometry and Auger electron spectroscopy (AES). Figure \ref{xrayV3} shows $\theta - 2\theta$ scans of TiN films grown on different substrates and buffer layers. Going from top to bottom, we see that on sapphire, growth on the bare substrate results in a mixture of (111)- and (200)-TiN, while the SiN buffer layer gives a film that is nearly all (200)-oriented, with a very weak (111) peak. The results from growth on H:i-Si at high temperature matches best with the SiN/sapphire, indicating a similar growth mode. The observed variation in the (200)-TiN diffraction peak position may reflect the difference in sample composition (especially the nitrogen content), the film deposition conditions( i.e temperature, the partial pressure of nitrogen and the power level), Ar and/or C incorporation into the lattice, and differences in the substrates.  See, for example, Figures 8 and 9 in J. -E. Sundgren et al. \cite{Sundgren}.

The (111)-TiN orientation is also observed when growing TiN on H:i-Si at low temperature, as shown in the bottom trace of Figure \ref{xrayV3}. The quality factors of resonators fabricated from these samples are significantly diminished relative to those made from films grown at high temperature, giving $Q_i$'s of 400,000 and 225,000 at high and low powers, respectively. This translates into losses from 2.5 - 4.5 $\times 10^{-6}$, comparable to those grown on sapphire. These data lead us to the hypothesis that, in the high temperature growth process, the silicon substrate is acquiring a layer of SiN that allows nucleation of the low-energy (200)-TiN growth. Comparing the low power data shown in Figure \ref{QvE}(b), we expect this layer to be relatively thin, about a factor of 10 less than the 50 nm buffer layer.  Furthermore, the very low loss in the single photon limit of the (200) TiN without a predeposited SiN buffer layer, $1.2\times10^6$, is an order of magnitude greater than measured in conventional superconducting resonators.  This low loss corresponds to a photon lifetime in excess of 10 microseconds.

We also found that the measured resonant frequencies of these devices is significantly lower than that expected by considering only the geometric inductance, suggesting significant kinetic inductance contribution. From the 40 nm thick, (200)-TiN films, we found the ratio $f_{r,meas}/f_{r,geom}=0.56\pm0.04$. On the other hand, the (111)-TiN grown on Si and sapphire had $f_{r,meas}/f_{r,geom}=0.30\pm0.01$. Using a variational method developed by Chang \cite{Chang, Mazin2}, we find the measured frequency shifts correspond to London penetration depths of $\lambda_L$= 275nm and 575nm, respectively. These numbers compare well with the penetration depths of 352 nm and 714 nm calculated from the measured resistance just above T$_C$ ($\rho_n$ = 45 $\mu\Omega$cm and 185 $\mu\Omega$ cm) using the BCS local relationship between $\lambda_L$, $\rho_n$, and the gap, $\Delta$ (i.e. $T_C$) \cite{Mattis,JGaoThesis}.

\begin{figure}[ht]
\centerline{\includegraphics[scale=.6]{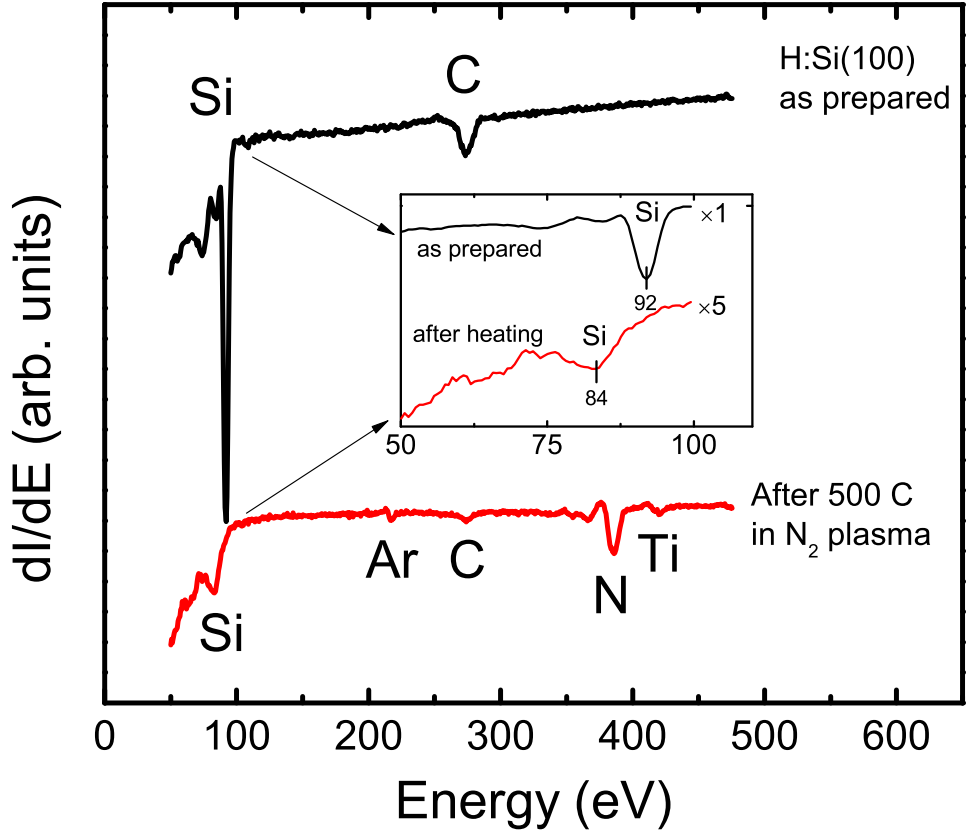}}
\caption{Auger spectra of the H:i-Si before and after the pre-deposition soak. The soak procedure consisited of the substrate being  raised to 500C, and the sputter gun being ramped up to the operational power in the working atmosphere of 3:2 Ar:N2. The inset shows a zoom onto the region of the Si(LMM) peak. The as-prepared substrate shows predominantly the free Si 92 eV peak with about 0.2 nm of the SiO, at 78 eV. After the soak, a SiN peak is observed at 84 eV. The small Ti and Ar peaks can be attributed to residual deposition around the shutter and implantation, respectively.}
\label{Auger}
\end{figure}

The {\it in situ} analysis also confirms that the Si in our high temperature growth process acquires a thin nitride layer prior to the TiN deposition. First, from ellipsometry we observe a small rotation of the light, corresponding to formation of a 1.5$\pm$0.5 nm layer of SiN. While this is near the lower limit of the resolution of an ellipsometer, these are very thick films for AES, which is extremely surface sensitive. In addition, the AES is element specific (probing depth $\approx$1 nm up to 400 eV) and gives chemical bonding information. The Auger spectra shown in Figure \ref{Auger} were taken from the Si substrate before and after the pre-sputtering soak.   Before the soak, the Si peak is at the unshifted energy of 92 eV, characteristic of a clean, unoxidized or nitrided surface. However, after the soak the Si peak has shifted down to 84 eV and a nitrogen peak of about the same size is in evidence. This shows that the surface is completely nitrided to at least 1 nm \cite{SiNAuger,SiOAuger}. These spectra show the presence of a  thin layer of SiN which is crucial for the formation of high quality TiN films. Finally, from the SEM analysis  we note that the SF$_6$ RIE used to pattern the TiN etches the Si at a higher rate, resulting in trenching around the CPW, leaving a flat undercut of $\approx 100$ nm under the TiN. This profile is further confirmation of a thin SiN layer under the TiN because it acts as a resist against the etch.

In conclusion, we find that the RF loss in TiN is dependent upon its growth mode, and is both substrate and temperature specific. The (200)-TiN growth results in resonators with internal $Q_i$ exceeding $10^7$ at high power.  The presence of (111)-TiN is correlated with depressed quality factors.  {\it In situ} analysis and tests using buffer layers showed that the (200)-TiN is stabilized on SiN. In the low power, single photon limit, quality factors up to $8\times10^5$ were observed from samples with the thinnest SiN layers.  High Q resonators fabricated from this low loss TiN have many applications, including quantum information and photon detectors.  Our result also opens the possibilities of making high Q TiN resonators on suspended SiN membranes, which may lead to other interesting applications.

\begin{acknowledgments} The authors would like to thank Ben Mazin and Jonas Zmuidzinas for the insightful discussions as well as Thomas Ohki and Chris Lirakis at BBN for guidance throughout the course of the work. CCT wishes to thank K. Saenger for useful discussions on XRD measurement. The views and conclusions contained in this document are those of the authors and should not be interpreted as representing the official policies, either expressly or implied, of the U.S. Government. \end{acknowledgments}

\newpage

\end{document}